\documentclass[conference]{IEEEtran}
\usepackage{amsmath, nccmath}
\usepackage{cases} 
\usepackage{amssymb}
\usepackage{cite}
\usepackage{url}
\usepackage{xcolor}
\usepackage{cite,graphicx}
\usepackage{subfigure}
\usepackage{citesort}
\usepackage{fancyhdr}
\usepackage{mdwmath}
\usepackage{mdwtab}
\usepackage{caption}
\usepackage{amsthm}
\usepackage{algorithmic}
\usepackage{algorithm}
\usepackage{threeparttable}
\usepackage{makecell}
\usepackage{booktabs}
\usepackage{pifont}

\newtheorem{remark}{Remark}
\newtheorem{theorem}{Theorem}

\newtheorem{lemma}{Lemma}

\newtheorem{corollary}{Corollary}

\makeatletter
\def\ScaleIfNeeded{%
\ifdim\Gin@nat@width>\linewidth \linewidth \else \Gin@nat@width
\fi } \makeatother

\begin{document}

\title{Pinching Antenna Systems (PASS) for Cognitive Radio (CR)}
\title{Beamforming Design for Pinching Antenna Systems Enabled Cognitive Radio Systems}

 \author{
\IEEEauthorblockN{ Haochen~Li\IEEEauthorrefmark{1}\IEEEauthorrefmark{2}, Ruikang~Zhong\IEEEauthorrefmark{3}, Zhaoming~Hu\IEEEauthorrefmark{4}, Zheng~Zhang\IEEEauthorrefmark{5}, Yaoyue~Hu\IEEEauthorrefmark{6}\IEEEauthorrefmark{2}, and Chao~Dong\IEEEauthorrefmark{1}} \IEEEauthorblockA{
\IEEEauthorrefmark{1} {College of Electronic and
Information Engineering, Nanjing University of Aeronautics and Astronautics, Nanjing, China}\\
\IEEEauthorrefmark{2} {National Mobile Communications Research Laboratory, Southeast University, Nanjing, China}\\
\IEEEauthorrefmark{3} {School of Information and Communication Engineering, Xi'an Jiaotong University, Xi'an, China}\\
\IEEEauthorrefmark{4} {College of Computer Science and Technology, Taiyuan University of Technology, Taiyuan, China}\\
\IEEEauthorrefmark{5} {School of Telecommunications Engineering, Xidian University, Xi’an, China}\\
\IEEEauthorrefmark{6} {School of Internet of Things, Nanjing University of Posts and Telecommunications, Nanjing, China}
 } }

\maketitle
\begin{abstract}
A pinching antenna system (PASS) assisted cognitive radio (CR) system is proposed. A secondary system sum rate maximization problem is formulated by jointly considering the base station (BS) power budget, the pinching antenna (PA) deployment constraints, and the interference tolerance requirements of primary users. To address the resulting non-convex problem, a tractable reformulation based on the weighted minimum mean-square error (WMMSE) approach is adopted, followed by the development of an alternating optimization (AO) algorithm. Within this framework, the auxiliary variables are updated in closed form, enabling an efficient transformation of the digital beamforming subproblem to a convex form, while the PA deployment is refined through a tailored element-wise optimization strategy. Numerical results validate the effectiveness of the proposed design and show consistent performance gains compared with conventional benchmark schemes.
\end{abstract}


\section{Introduction}\label{I}
The explosive growth of wireless services has led to an increasing demand for radio spectrum. However, the conventional static spectrum allocation policy results in significant underutilization of licensed frequency bands, thereby intensifying the spectrum scarcity~\cite{8766143}. To overcome this limitation, cognitive radio (CR) has emerged as a pivotal technology that enables the efficient reuse of licensed spectrum resources. In underlay CR systems, secondary users are allowed to share the spectrum with primary users while ensuring that the interference imposed on the latter remains below a tolerable threshold. Owing to its advantages of reducing the dependence on accurate spectrum sensing and simultaneous spectrum utilization by both primary and secondary users, the underlay mode CR is widely investigated~\cite{5783948}. 
Interference management via beamforming design constitutes a key research topic. Nevertheless, the performance gains achievable through beamforming in CR systems become marginal under certain circumstances: (i) when the channel between the base station (BS) and the secondary user is weak, or (ii) when the channel between the BS and the primary user is strong. In both cases, the BS faces unfavorable channel conditions, which prevent it from delivering sufficient signal power to secondary users without simultaneously causing notable interference to primary users~\cite{5639025}. To tackle these challenges, pinching antenna systems(PASS) have been introduced into CR systems. PASS deploys elevated waveguides and applies antenna elements called pinching antennas (PAs) along the waveguide, thereby enabling tunable and reconfigurable wireless channels~\cite{11169486}. This unique design allows the BS to flexibly adjust antenna deployment without requiring physical relocation or complex hardware reconfiguration~\cite{11364174}.  The PASS enabled CR system has been preliminarily investigated in~\cite{sun2025pinching}, demonstrating the promising potential of PASS for CR applications. However, this study is restricted to a highly simplified system configuration involving only a single primary user and a single secondary user, which falls short of capturing the complexity of practical deployment scenarios. Furthermore, the adoption of a single waveguide at the BS inherently limits its spatial multiplexing capability, thereby constraining system performance. To date, the beamforming design for general multi-waveguide PASS-enabled multi-user CR systems remains largely underexplored, calling for more comprehensive investigation. To fill this research gap, this work focus on integrating PASS into CR system to facilitate signal beamforming. The main contributions of this work can be summarized as follows:
\begin{itemize}
        \item A PASS assisted CR system is proposed, where the PASS is leveraged to enhance the secondary system capacity by improving the BS–secondary user channel gain while suppressing the BS–primary user interference. A secondary system capacity maximization problem is formulated by jointly optimizing the digital beamforming at the BS and the pinching beamforming at the PASS under BS power constraint, PA deployment constraint, and primary user interference constraint. 
        \item For the \emph{general case} with multiple primary/secondary users, multiple waveguides and multiple PAs per waveguide, the optimization problem is first reformulated using the weighted minimum mean-square error (WMMSE) method. An alternating optimization (AO) framework is then developed to iteratively optimize the auxiliary variables, the digital beamforming matrix, and the PA deployment matrix. The auxiliary variables admit closed-form solutions, which transform the digital beamforming problem into a convex form. The pinching beamforming problem is tackled via an element-wise optimization method.
\end{itemize}

\section{System Model}\label{II}
\begin{figure} [htbp]
\centering\vspace{-0.2cm}
\includegraphics[width=0.4\textwidth]{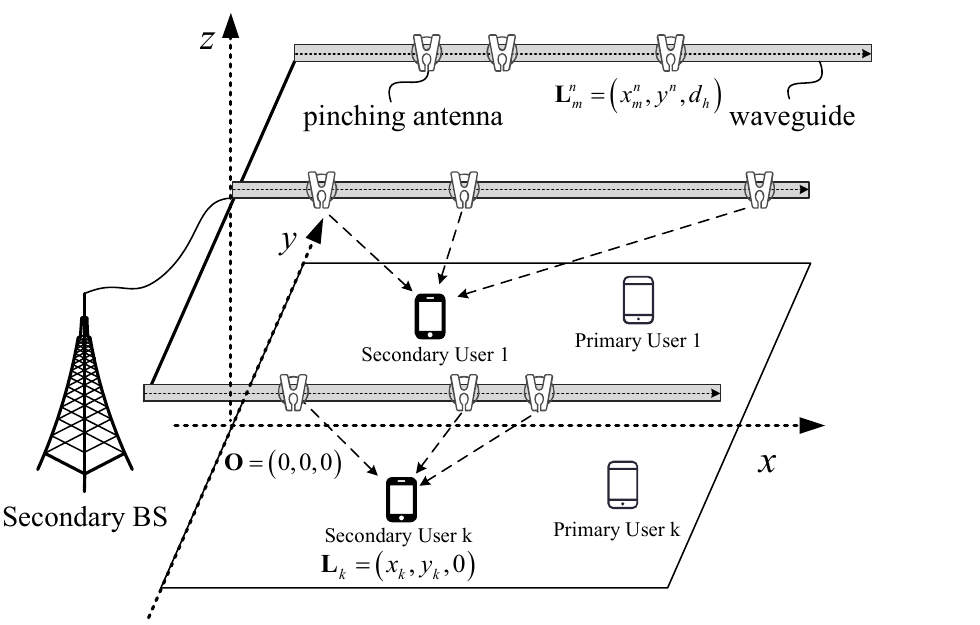}
 \caption{The proposed PASS assisted CR system.
  }
 \label{system_model}\vspace{-0.2cm}
\end{figure}

As depicted in Fig.~\ref{system_model}, a PASS assisted CR system is investigated in this work, where the CR BS reuses the spectrum of $P$ primary users to serve $K$ secondary users through pinching antennas.  The CR BS adopts the PASS architecture with $N$ waveguides and $M$ PAs per waveguide. The set of all waveguides, all PAs on a waveguides, all primary users, and all secondary users are   $\mathcal{N}=\left[1,2,\cdots,N\right]$, $\mathcal{M}=\left[1,2,\cdots,M\right]$,  $\mathcal{P}=\left[1,2,\cdots,P\right]$, and $\mathcal{K}=\left[1,2,\cdots,K\right]$, respectively. 

The service area, which contains all primary and secondary users, is defined as a rectangular region with size $A = D_x \times D_y$. The location of a primary user $p$ is denoted by $\boldsymbol{l}_p = [x_p, y_p, 0]$, $\forall p \in \mathcal{P}$, while the location of a secondary user $k$ is denoted by $\boldsymbol{l}_k = [x_k, y_k, 0]$, $\forall k \in \mathcal{K}$. The waveguides of the PASS are deployed above the service area and extend parallel to the $x$-axis. The feed point of the $n$-th waveguide is located at $\boldsymbol{l}_n = [0, y_n, d_h]$, $\forall n \in \mathcal{N}$, where $d_h$ is the height of PASS. Moreover, the position of the $m$-th PA attached to the $n$-th waveguide is given by $\boldsymbol{l}_n^m = [x_n^m, y_n, d_h]$, $\forall n \in \mathcal{N}, \forall m \in \mathcal{M}$. The PASS system can achieve pinching beamforming via adjusting the attached location of PAs on waveguides, i.e., tuning $ x_n^m, \forall n \in \mathcal{N}, \forall m \in \mathcal{M}$. Suppose the length of waveguides in PASS is $L$, and the minimum distance between adjacent PAs is $\delta=\lambda_c/2$, where $\lambda_c$ is the carrier wavelength. The location $x_n^m$
of the $m$-th PA attached to the $n$-th waveguide is required to satisfy the following constraint
\begin{equation}\label{X}
\begin{aligned}
\mathcal{X}=\Big\{ {\begin{array}{*{20}{c}}
{x_n^m,}&\vline& {0 \le x_n^m \le L}\\
{ \forall n, m}&\vline& {x_n^m - x_n^{m-1} \ge \delta, \  \text{if} \  m>1}
\end{array}} \Big\}.
\end{aligned}
\end{equation}

\subsection{Channel Model}
The channel between the PASS and users can be decomposed into two components: the wireless channel and the in-waveguide channel. The wireless channel between the $n$-th waveguide in the PASS and secondary user $k$ or primary user $p$ can be expressed as
\begin{equation}
        \mathbf{h}_{k,n}\!=\!\!\left(2\kappa_c\right)^{-1}\!\Big[\frac{e^{-j \kappa_c r_{k,n}^1}} {r_{k,n}^1},\frac{e^{-j \kappa_c r_{k,n}^2}} {r_{k,n}^2},\cdots,\frac{e^{-j \kappa_c r_{k,n}^M}} {r_{k,n}^M}\Big]^\mathrm{T}
\end{equation}
\vspace{0.05cm}\begin{equation}
        \mathbf{g}_{p,n}\!=\!\!\left(2\kappa_c\right)^{-1}\!\Big[\frac{e^{-j \kappa_c r_{p,n}^1}} {r_{p,n}^1},\frac{e^{-j \kappa_c r_{p,n}^2}} {r_{p,n}^2},\cdots,\frac{e^{-j \kappa_c r_{p,n}^M}} {r_{p,n}^M}\Big]^\mathrm{T}
\end{equation}
where $\kappa_c=2\pi/\lambda_c$ denotes the wave number. Specifically, $r_{k,n}^m=\left\|\boldsymbol{l}_n^m-\boldsymbol{l}_k\right\|_2$ and $r_{p,n}^m=\left\|\boldsymbol{l}_n^m-\boldsymbol{l}_p\right\|_2$ represent the wireless propagation distances between the $m$-th PA on the $n$-th waveguide and the secondary user $k$, and between the $m$-th PA on the $n$-th waveguide and the primary user $p$, respectively. By stacking the channel vectors corresponding multiple waveguides, the channel between the $n$-th waveguide in the PASS and secondary user $k$ or primary user $p$ can be expressed as
\begin{equation}\label{wireless_k}
       \hat{\mathbf{h}}_k=\left[\mathbf{h}_{k,1}^\mathrm{T},\mathbf{h}_{k,2}^\mathrm{T},\cdots,\mathbf{h}_{k,N}^\mathrm{T}\right]^\mathrm{T}\in\mathbb{C}^{MN\times 1}, \forall k \in \mathcal{K},
\end{equation}
\begin{equation}\label{wireless_p}
       \hat{\mathbf{g}}_p=\left[\mathbf{g}_{p,1}^\mathrm{T},\mathbf{g}_{p,2}^\mathrm{T},\cdots,\mathbf{g}_{p,N}^\mathrm{T}\right]^\mathrm{T}\in\mathbb{C}^{MN\times 1}, \forall p \in \mathcal{P}. 
\end{equation}
The in-waveguide describe the signal propagation between the feed point of the waveguide and the PA. Specifically, for the $m$-th PA attached to the $n$-th waveguide, the corresponding in-waveguide channel element can be expressed as 
\begin{equation}
        f_n^m = \rho_n^me^{-j\kappa_gd_n^m},        
\end{equation}
where $\kappa_g=n_e\kappa_c$ is the guided wave number given waveguide  effective refractive index $n_e$. $d_n^m=\left\|\boldsymbol{l}_n^m-\boldsymbol{l}_n\right\|_2$ is the in-waveguide propagation distance between the $m$-th PA on the $n$-th waveguide and its feed point. $\rho_n^m$ is the power allocation factor for PAs on the $n$-th waveguide, satisfying $\sum\nolimits_{m}\left(\rho_n^m\right)^2=1$. The even power allocation strategy is adopts in this work, which yields $\rho_n^m=1/\sqrt{M}, \forall m,n$. The in-waveguide channel vector between the PAs on the $n$-th waveguide and its feed point can be expressed as
\begin{equation}
        \mathbf{f}_n=\left[f^1_n,f^2_n,\cdots,f^M_n\right]^\mathrm{T}\in\mathbb{C}^{M\times 1}.
\end{equation}
As each PA can only be attached to one specific waveguide, the overall in-waveguide channel of the PASS is in the form of block diagonal matrix, which can be written as
\begin{equation}\label{in-waveguide}
\begin{aligned}
\mathbf{F}=\text{blkdiag}\left(\mathbf{f}_1,\mathbf{f}_2,\cdots,\mathbf{f}_N\right)^\mathrm{T}\in\mathbb{C}^{N\times MN}
\end{aligned}
\end{equation}
Combining~\eqref{wireless_k},~\eqref{wireless_p} and~\eqref{in-waveguide}, the PASS channel for secondary user $k$ or primary user $p$  can be expressed as 
\begin{equation}\label{h_k}
        \mathbf{h}_k=\mathbf{F}\hat{\mathbf{h}}_k\in\mathbb{C}^{N\times 1},  \forall k \in \mathcal{K},
\end{equation}
\begin{equation}\label{g_p}
        \mathbf{g}_p=\mathbf{F}\hat{\mathbf{g}}_p\in\mathbb{C}^{N\times 1}, \forall p \in \mathcal{P}.
\end{equation}
\subsection{Signal Model}
The signal received by the secondary user $k$ can be expressed as 
\begin{equation}
\begin{aligned}
        y_k = \mathbf{h}_k^\mathrm{H}\mathbf{W}\mathbf{s}+n_k
        = \mathbf{h}_k^\mathrm{H}\mathbf{w}_ks_k+\!\sum\nolimits_{j\ne k}\!\mathbf{h}_k^\mathrm{H}\mathbf{w}_js_j+n_k,
\end{aligned}
\end{equation}
where $\mathbf{W}=\left[\mathbf{w}_1,\mathbf{w}_2,\cdots,\mathbf{w}_K\right]\in\mathbb{C}^{N\times K}$ is the beamforming matrix with $\mathbf{w}_k$ denoting the beamforming vector for secondary user $k$. $\mathbf{s}=\left[{s}_1,{s}_2,\cdots,{s}_K\right]^\mathrm{T}\in\mathbb{C}^{K\times 1}$ is the signal vector with ${s}_k$ denoting the information baring signal for secondary user $k$. The stochastic signal is normalized and uncorrelated between different users, i.e., $\mathbb{E}\left\{\mathbf{s} \mathbf{s} ^\mathrm{H}\right\}=\mathbf{I}_K$. $n_k\sim\mathcal{CN}(0, \sigma^2)$ stands for the additive white Gaussian noise~(AWGN) at the receiver of user $k$ with $\sigma^2$ denoting the noise power. The signal-to-interference-plus-noise ratio~(SINR) of secondary user $k$ can be expressed as
\begin{equation}
        \text{SINR}_k=\frac{\left|\mathbf{h}_{k}^\mathrm{H}\mathbf{w}_{k}\right|^2}{\sum_{j\neq k}\left|\mathbf{h}_{k}^\mathrm{H}\mathbf{w}_{j}\right|^2+\sigma^2}.
\end{equation}
The communication rate of secondary user $k$ is 
\begin{equation}\label{R_k}
        R_k = \ln\left(\text{SINR}_k+1\right).
\end{equation}
The interference power caused by PASS at primary user $p$, which termed as interference temperature (IT) in CR system can be expressed as
\begin{equation}
       \text{IT}_p = \left\|\mathbf{W}^\mathrm{H}\mathbf{g}_p\right\|^2_2=\sum\nolimits_j\left|\mathbf{g}_p^\mathrm{H}\mathbf{w}_j\right|^2.
\end{equation}

\subsection{Problem Formulation}
The objective of this work is to maximize the sum rate of secondary users, subject to limited computational resources at the BS and controlled interference at all primary users. The corresponding optimization problem is formulated as
\begin{subequations}\label{problem:sum_rate}
    \begin{align}        
        \max_{\mathbf{W},\mathbf{X}} \quad &  {\sum\nolimits_{k\in\mathcal{K}}\ln\left(1+\text{SINR}_k\right)} \\
        \label{constraint:PASS}
        \mathrm{s.t.} \quad & \left[\mathbf{X}\right]_{mn} \in \mathcal{X}, \forall m \in \mathcal{M},n\in \mathcal{N},  \\ 
        \label{constraint:power}
        & \text{tr}\left(\mathbf{W}^\mathrm{H}\mathbf{W}\right)\le P_b,\\
        \label{constraint:IT}
        & \text{IT}_p\le \gamma_p, \forall p \in \mathcal{P},
    \end{align}
\end{subequations}
where $\mathbf{X}$ denotes the PA deployment matrix, with its $(m,n)$-th entry given by $x_n^m$. Constraint~\eqref{constraint:PASS} specifies the feasible PA deployment region, which is determined by the minimum PA spacing $\delta$ and the waveguide length $L$. Constraint~\eqref{constraint:power} imposes the power budget $P_b$ at the CR BS. Constraint~\eqref{constraint:IT} ensures that the interference introduced to each primary user~$p$ remains below the predefined threshold $\gamma_p$.

\section{Proposed Alternating Optimization Algorithm for Multiple  Secondary User}\label{multiple}
\subsection{Reformulation of Problem~\eqref{problem:sum_rate}}
First, the lemma used for reformulate problem~\eqref{problem:sum_rate} is  introduced.

\begin{lemma}[WMMSE~\cite{shi2015secure}]
\emph{Let $\mathbf{c}\in\mathbb{C}^{N\times 1}$, $\mathbf{d}\in\mathbb{C}^{N\times 1}$, $\alpha\in\mathbb{C}$, and $e\in\mathbb{R}$ satisfy $\mathbf{c}^\mathrm{H}\mathbf{d}\neq 0$ and $e>0$. Define 
\begin{align*}
    z\left(\alpha,e,\mathbf{c},\mathbf{d}\right) &= \left|1-\alpha^*\mathbf{c}^\mathrm{H}\mathbf{d}\right|^2+\left|\alpha\right|^2e \\
              &= \left(1-\alpha^*\mathbf{c}^\mathrm{H}\mathbf{d}\right)\left(1-\alpha^*\mathbf{c}^\mathrm{H}\mathbf{d}\right)^*+\alpha\alpha^* e,
\end{align*}
\[
    g\left(\eta,\alpha,e,\mathbf{c},\mathbf{d}\right)=\ln(\eta)-\eta z \left(\alpha,e,\mathbf{c},\mathbf{d}\right).
\]
Then, the following equality holds:}
\begin{equation}\label{reformulate}
        \ln\left(1+e^{-1}|\mathbf{c}^\mathrm{H}\mathbf{d}|^2\right)=\max_{\eta>0,\alpha}g\left(\eta,\alpha,e,\mathbf{c},\mathbf{d}\right)
\end{equation}

\end{lemma}
Using Lemma 1 to~\eqref{R_k}, it can be reformulated to 
\begin{equation}
\begin{aligned}
       R_k&=\ln\Big(1+\big({\sum\nolimits_{j\neq k}\left|\mathbf{h}_{k}^\mathrm{H}\mathbf{w}_{j}\right|^2+\sigma^2}\big)^{-1}{\left|\mathbf{h}_{k}^\mathrm{H}\mathbf{w}_{k}\right|^2}\Big)\\
       =&\max_{\eta_k>0,\alpha_k}g\Big(\eta_k,\alpha_k,\sum\nolimits_{j\neq k}\left|\mathbf{h}_{k}^\mathrm{H}\mathbf{w}_{j}\right|^2+\sigma^2,\mathbf{h}_k,\mathbf{w}_k\Big).
\end{aligned} 
\end{equation}
Based on this reformulation, problem~\eqref{problem:sum_rate} can be reformulated as 
\begin{subequations}\label{problem:sum_rate_reformulated}
    \begin{align}        
        \max_{\mathbf{W},\mathbf{X},\{\alpha_k,\eta_k>0\}_{k=1}^K} \  &  {\sum_{k\in\mathcal{K}}g\Big(\eta_k,\alpha_k,\sum_{j\neq k}\left|\mathbf{h}_{k}^\mathrm{H}\mathbf{w}_{j}\right|^2+\sigma^2,\mathbf{h}_k,\mathbf{w}_k\Big)} \\       
        \mathrm{s.t.} \quad & \eqref{constraint:PASS},\eqref{constraint:power},\eqref{constraint:IT}.
    \end{align}
\end{subequations}
Problem~\eqref{problem:sum_rate_reformulated} is addressed using an alternating optimization framework.  To facilitate the solution, all variables are partitioned into four parts: $\{\alpha_k\}_{k=1}^K$, $\{\eta_k\}_{k=1}^K$, $\mathbf{W}$, and $\mathbf{X}$. At each iteration, the variables in one part are optimized while those in the other parts are kept fixed.
\subsection{Optimizing auxiliary variables and Solving Digital Beamforming Problem}\label{digital}
With fixed $\{\eta_k\}_{k=1}^K$, $\mathbf{W}$, and $\mathbf{X}$, the optimal solution to $\alpha_k$ can be obtained applying the first-order optimality condition
\begin{equation}\label{alpha_opt_k}
       \alpha_k^\text{opt}=\Big(\sum\nolimits_{j}\left|\mathbf{h}_{k}^\mathrm{H}\mathbf{w}_{j}\right|^2+\sigma^2\Big)^{-1}\mathbf{h}_{k}^\mathrm{H}\mathbf{w}_{k}.
\end{equation}
With fixed $\{\alpha_k\}_{k=1}^K$, $\mathbf{W}$, and $\mathbf{X}$, the optimal solution to $\eta_k$ can be obtained applying the first-order optimality condition
\begin{equation}\label{eta_opt_k}
       \eta_k^\text{opt}={z\Big(\alpha_k,\sum\nolimits_{j\neq k}\left|\mathbf{h}_{k}^\mathrm{H}\mathbf{w}_{j}\right|^2+\sigma^2,\mathbf{h}_{k},\mathbf{w}_{k}\Big)}^{-1}.
\end{equation}

With fixed auxiliary variables $\{\alpha_k\}_{k=1}^K$, $\{\eta_k\}_{k=1}^K$, and PA deployment setting, the digital beamforming problem can be expressed as 
\begin{subequations}\label{problem:sum_rate_W}
    \begin{align}        
        \min_{\mathbf{W}} \quad &  \sum\nolimits_{k} \mathbf{w}_k^\mathrm{H}\mathbf{C}\mathbf{w}_k-2\sum\nolimits_{k}\Re\left(\mathbf{u}_{k}^\mathrm{H}\mathbf{w}_{k}\right)\\
        \mathrm{s.t.} \quad & \eqref{constraint:power}, \eqref{constraint:IT},
    \end{align}
\end{subequations}
where $\mathbf{C}=\sum_{j}\eta_j\left|\alpha_j\right|^2\mathbf{h}_{j}^\mathrm{H}\mathbf{h}_{j}$, $\mathbf{u}_{k}=\eta_k\alpha_k\mathbf{h}_{k}$. Problem~\eqref{problem:sum_rate_W} is convex and can be effectively solved with CVX. 
\subsection{Pinching Beamforming Problem}\label{pinching}
With fixed auxiliary variables $\{\alpha_k\}_{k=1}^K$, $\{\eta_k\}_{k=1}^K$, and digital beamforming design, the pinching beamforming problem can be expressed as 
\begin{subequations}\label{problem:sum_rate_X}
    \begin{align}        
        \min_{\mathbf{X}} \quad &  \sum\nolimits_{k} \mathbf{h}_k^\mathrm{H}\mathbf{D}_k\mathbf{h}_k-2\sum\nolimits_{k}\Re\left(\mathbf{v}_{k}^\mathrm{H}\mathbf{h}_{k}\right) \\
        \mathrm{s.t.} \quad & \eqref{constraint:PASS}, \eqref{constraint:IT}.
    \end{align}
\end{subequations}
where $\mathbf{D}_k=\eta_k\left|\alpha_j\right|^2\sum_j\mathbf{w}_j\mathbf{w}_j^\mathrm{H}$, $\mathbf{v}_{k}=\eta_k\alpha_k^*\mathbf{w}_{k}$. The channel between the $m$-th PAs on all waveguides and all $K$ secondary user is 
\begin{equation}
        \mathbf{H}^m  = \left[\mathbf{h}_1^m ,\mathbf{h}_2^m ,\cdots,\mathbf{h}_K^m \right],
\end{equation}
where $\mathbf{h}_k^m$ is the channel between the $m$-th PAs on all waveguides and secondary user $k$. 
Then, the PASS channel for secondary
user $k$ in~\eqref{h_k} can be rewritten as
\begin{equation}\label{h_k1}
\begin{aligned}
        &\mathbf{h}_k=\left[\mathbf{f}_1^\mathrm{T}\mathbf{h}_{k,1},\mathbf{f}_2^\mathrm{T}\mathbf{h}_{k,2},\cdots,\mathbf{f}_N^\mathrm{T}\mathbf{h}_{k,N} \right]^\mathrm{T}=\\ &\Big[\sum\nolimits_m{f}_1^m[\mathbf{h}_{k,1}]_m,\cdots,\!\sum\nolimits_m{f}_N^m[\mathbf{h}_{k,N}]_m\Big]^\mathrm{T} \!\!\!\!\!=\! \sum\nolimits_m\mathbf{h}_k^m,
\end{aligned}
\end{equation}
The channel between the PASS and all primary user is
\begin{equation}
        \mathbf{H}  = \left[\mathbf{h}_1 ,\mathbf{h}_2 ,\cdots,\mathbf{h}_K \right]=\sum\nolimits_m\mathbf{H}^m.
\end{equation}
Similarly, the PASS channel for primary user $p$ in~\eqref{g_p} can be rewritten as
\begin{equation}\label{g_p1}
\begin{aligned}
        &\mathbf{g}_p=\left[\mathbf{f}_1^\mathrm{T}\mathbf{g}_{p,1},\mathbf{f}_2^\mathrm{T}\mathbf{g}_{p,2},\cdots,\mathbf{f}_N^\mathrm{T}\mathbf{g}_{p,N}  \right]^\mathrm{T}=\\ &\Big[\sum\nolimits_m{f}_1^m[\mathbf{g}_{p,1}]_m,\cdots,\!\sum\nolimits_m{f}_N^m[\mathbf{g}_{p,N}]_m\Big]^\mathrm{T}\!\!\!\!=\! \sum\nolimits_m\mathbf{g}_p^m,
\end{aligned}
\end{equation}
where $\mathbf{g}_p^m$ is the channel between the $m$-th PAs on all waveguides and primary user $p$. The channel between the $m$-th PAs on all waveguides and all primary users is 
\begin{equation}
        \mathbf{G}^m  = \left[\mathbf{g}_1^m ,\mathbf{g}_2^m ,\cdots,\mathbf{g}_P^m \right].
\end{equation}
The channel between the PASS and all primary user is
\begin{equation}
        \mathbf{G}  = \left[\mathbf{g}_1 ,\mathbf{g}_2 ,\cdots,\mathbf{g}_P \right]=\sum_m\mathbf{G}^m.
\end{equation}
Define auxiliary matrices and vectors
\begin{equation}\label{auxiliary_T}
        \mathbf{T}=\left[\mathbf{t}_1,\mathbf{t}_2,\cdots,\mathbf{t}_K\right] = \sum_m\mathbf{T}^m.
\end{equation}
\begin{equation}\label{auxiliary_Q}
        \mathbf{Q}=\left[\mathbf{q}_1,\mathbf{q}_2,\cdots,\mathbf{q}_P\right] = \sum_m\mathbf{Q}^m,
\end{equation}
With these auxiliary matrices, problem~\eqref{problem:sum_rate_X} can be reformulated as 
\begin{subequations}\label{problem:sum_rate_X_Re}
    \begin{align}        
        \min_{\mathcal{A}} \quad &  \sum\nolimits_{k} \mathbf{t}_k^\mathrm{H}\mathbf{D}_k\mathbf{t}_k-2\sum\nolimits_{k}\Re\left(\mathbf{v}_{k}^\mathrm{H}\mathbf{t}_{k}\right) \\
        \mathrm{s.t.} \quad & \mathbf{T}^m=\mathbf{H}^m, \mathbf{Q}^m=\mathbf{G}^m, \forall m,\\ 
        & \mathbf{T}=\sum\nolimits_m\mathbf{T}^m, \mathbf{Q}=\sum\nolimits_m\mathbf{Q}^m,\\ 
        \label{constraint:IT_Q}
        & \left\|\mathbf{W}^\mathrm{H}\mathbf{q}_p\right\|_2^2 \le {\gamma}_p, \forall p, \eqref{constraint:PASS},
    \end{align}
\end{subequations}
where $\mathcal{A}=\Big\{\mathbf{X},\mathbf{Q},\left\{\mathbf{Q}^m\right\}_{m=1}^{M},\mathbf{T},\left\{\mathbf{T}^m\right\}_{m=1}^{M}\Big\}$ denotes the set containing all optimization variables. The penalty  method is introduced to handle the auxiliary equality constraints in problem~\eqref{problem:sum_rate_X_Re}. The optimization problem incorporating  penalty terms in the objective function can be expressed as
\begin{subequations}\label{problem:sum_rate_X_Re_penalty}
    \begin{align}        
        \min_{\mathcal{A}} \quad &  g\left(\mathcal{A}\right) \\
        \mathrm{s.t.} \quad  & \eqref{constraint:PASS},\eqref{constraint:IT_Q} ,
    \end{align}
\end{subequations}
where the objective function of problem~\eqref{problem:sum_rate_X_Re_penalty} is given as~\eqref{g_penalty} at the top of the next page. $\tau$ denotes the penalty factor.
\begin{figure*}[!t]
\normalsize\vspace{0.1cm}
\begin{equation}\label{g_penalty}
\begin{aligned}
     \!\!\!g\left(\mathcal{A}\right)\!\!=\!\!\sum_{k} \mathbf{t}_k^\mathrm{H}\mathbf{D}_k\mathbf{t}_k\!\!-2\!\!\sum_{k}\!\Re\left(\mathbf{v}_{k}^\mathrm{H}\mathbf{t}_{k}\right)\!+\!\frac{1}{\tau}\Big\|\mathbf{T}-\!\!\sum_{m=1}^M\!\!\mathbf{T}^m\Big\|_F^2\!\!\!+\!\frac{1}{\tau} \!\sum_{m=1}^M\!\!\left\|\mathbf{T}^m\!-\!\mathbf{H}^m\right\|_F^2\!\!+\!\frac{1}{\tau}\Big\|\mathbf{Q}-\!\!\sum_{m=1}^M\!\!\mathbf{Q}^m\Big\|_F^2\!\!\!+\!\frac{1}{\tau} \!\sum_{m=1}^M\!\!\left\|\mathbf{Q}^m-\mathbf{G}^m\right\|_F^2.
\end{aligned}
\end{equation}
\hrulefill \vspace*{0pt}
\end{figure*}
Problem~\eqref{problem:sum_rate_X_Re_penalty} is addressed using an AO framework. To facilitate the solution, all variables are partitioned into four parts: $\mathbf{Q}$, $\mathbf{T}$, $\big\{\{\mathbf{Q}^m\}_{m=1}^{M},\{\mathbf{T}^m\}_{m=1}^{M}\big\}$, and $\mathbf{X}$.
\subsubsection{Optimizing $\mathbf{Q}$ }With given $\mathbf{T}$, \!$\big\{\{\mathbf{Q}^m\}_{m=1}^{M},\{\mathbf{T}^m\}_{m=1}^{M}\big\}$, and $\mathbf{X}$ the optimization problem with respect to $\mathbf{Q}$ can be expressed as
\begin{subequations}\label{problem:sum_rate_X_Re_penalty_Q}
    \begin{align}        
        \min_\mathbf{Q} \quad &  \Big\|\mathbf{Q}-\sum_{m=1}^M\mathbf{Q}^m\Big\|_F^2 \\
        \mathrm{s.t.} \quad  & \eqref{constraint:IT_Q} .
    \end{align}
\end{subequations}
Problem~\eqref{problem:sum_rate_X_Re_penalty_Q} can be decomposed into $P$ independent optimization problems. The optimization problem with respect to $\mathbf{q}_p$ can be expressed as
\begin{subequations}\label{problem:sum_rate_X_Re_penalty_Q_p}
    \begin{align}        
        \min_{ \mathbf{q}_p} \quad &  \left\|\mathbf{q}_p-\mathbf{a}_p\right\|_2^2 \\
        \mathrm{s.t.} \quad  & \mathbf{q}_p^\mathrm{H}\mathbf{W}\mathbf{W}^\mathrm{H}\mathbf{q}_p \le {\gamma}_p.
    \end{align}
\end{subequations}
 Problem~\eqref{problem:sum_rate_X_Re_penalty_Q_p} is convex and following proposition  gives the closed form solution to problem~\eqref{problem:sum_rate_X_Re_penalty_Q_p}. Using the Lagrangian method, the closed form solution to problem~\eqref{problem:sum_rate_X_Re_penalty_Q_p} can be given by
\begin{equation}\label{tilde_solution}
    \tilde{\mathbf q}_p^\text{opt} =
    \left(\mathbf{I}_N+\tilde{\mu}_p\mathbf{W}\mathbf{W}^\mathrm{H}\right)^{-1} \mathbf{a}_p.
\end{equation}
The Lagrangian multiplier can be expressed as
\begin{equation}
    \tilde{\mu}_p =
    \begin{cases}
        0,\!\!& \|\mathbf{W}^\mathrm{H}\mathbf{a}_p\|_2^2 \le {\gamma}_p, \\
        \displaystyle \!\arg\left\{ \tilde{\mu}_p \ge 0 \ \middle|
    \tilde{g}_p\left(\tilde{\mu}_p\right) \!=\! \gamma_p \right\},\!\!& \|\mathbf{W}^\mathrm{H}\mathbf{a}_p\|_2^2 > {\gamma}_p.
    \end{cases}
\end{equation}

\subsubsection{Optimizing $\mathbf{T}$ }  With given $\mathbf{Q}$, \!$\big\{\{\mathbf{Q}^m\}_{m=1}^{M},\{\mathbf{T}^m\}_{m=1}^{M}\big\}$, and $\mathbf{X}$, the optimization problem with respect to $\mathbf{T}$ can be expressed as
\begin{equation}\label{problem:sum_rate_X_Re_penalty_T}    
        \!\min_\mathbf{T} \ \sum_{k} \mathbf{t}_k^\mathrm{H}\mathbf{D}_k\mathbf{t}_k\!-\!2\sum_{k}\Re\left(\mathbf{v}_{k}^\mathrm{H}\mathbf{t}_{k}\right)\!+\!\frac{1}{\tau}\Big\|\mathbf{T}\!-\!\sum_{m=1}^M\!\mathbf{T}^m\Big\|_F^2.
\end{equation}
Based on the stationarity condition, the optimal solution to problem~\eqref{problem:sum_rate_X_Re_penalty_T} can be expressed as
\begin{equation}\label{opt_T}
\mathbf{t}_k^\text{opt}=\Big(\frac{1}{\tau}\mathbf{I}_N+\mathbf{D}_k\Big)^{-1}\Big(\mathbf{v}_k+\frac{1}{\tau}\Big[\sum\nolimits_{m=1}^M\mathbf{T}^m\Big]_{:,k}\Big) .
\end{equation}
\subsubsection{Optimizing $\big\{\{\mathbf{Q}^m\}_{m=1}^{M},\{\mathbf{q}^m\}_{m=1}^{M}\big\}$} 
With given $\mathbf{Q}$, $\mathbf{T}$, and $\mathbf{X}$, the optimization problem with respect to $\big\{\{\mathbf{Q}^m\}_{m=1}^{M},\{\mathbf{T}^m\}_{m=1}^{M}\big\}$ can be expressed as
\begin{equation}\label{problem:sum_rate_X_Re_penalty_QmTm} \min_{\{\mathbf{Q}^m,\mathbf{q}^m\}_{m=1}^{M}} \quad  \tilde{f}\left(\{\mathbf{Q}^m\}_{m=1}^{M},\{\mathbf{T}^m\}_{m=1}^{M}\right), 
\end{equation}
where the objective function of problem~\eqref{problem:sum_rate_X_Re_penalty_QmTm} is given as~\eqref{tilde_g_penalty} at the top of the next page. 
\begin{figure*}[!t]
\normalsize
\begin{equation}\label{tilde_g_penalty}
\begin{aligned}
     \!\!\tilde{g}\left(\{\mathbf{Q}^m\}_{m=1}^{M},\{\mathbf{T}^m\}_{m=1}^{M}\right)\!\!=\!\!\Big\|\mathbf{T}-\!\sum\nolimits_{m=1}^M\!\!\mathbf{T}^m\Big\|_F^2\!\!+ \!\sum\nolimits_{m=1}^M\!\!\left\|\mathbf{T}^m\!-\!\mathbf{H}^m\right\|_F^2+\Big\|\mathbf{Q}-\!\sum\nolimits_{m=1}^M\!\!\mathbf{Q}^m\Big\|_F^2\!\!+ \!\sum\nolimits_{m=1}^M\!\!\left\|\mathbf{Q}^m-\mathbf{G}^m\right\|_F^2.
\end{aligned}
\end{equation}
\hrulefill \vspace*{0pt}
\end{figure*}
Based on the stationarity condition, solutions to problem~\eqref{problem:sum_rate_X_Re_penalty_QmTm} are
\begin{equation}\label{opt_Qm}
        (\mathbf{Q}^{m})^\text{opt}=\frac{1}{M+1}\Big(\mathbf{Q}-\sum_{i=m}^M\mathbf{G}^i\Big)+\mathbf{G}^m, \forall m,
\end{equation}
\begin{equation}\label{opt_Tm}
        (\mathbf{T}^{m})^\text{opt}=\frac{1}{M+1}\Big(\mathbf{T}-\sum_{i=m}^M\mathbf{H}^i\Big)+\mathbf{H}^m, \forall m.
\end{equation}
\begin{algorithm}[t]
\caption{Penalty algorithm for solving problem~\eqref{problem:sum_rate_X_Re_penalty}.}\label{algorithm4}
\begin{algorithmic}[1]
\STATE {Initialize feasible $\mathbf{T}^{(0)}$, $\mathbf{T}_m^{(0)}$, and $\mathbf{Q}_m^{(0)}$ for all $m$, PA deployment variable set $\mathbf{X}^{(0)}$, the AO convergence tolerance $\epsilon_1 > 0$, the penalty iteration convergence tolerance $\epsilon_2 > 0$, the penalty factor $\tau$,  and the penalty update factor $\omega<1$. Set the iteration index  $t=0$.}
\REPEAT
\REPEAT
\STATE Given $\mathbf{T}^{(t)}$, $\mathbf{T}_m^{(t)}$, $\mathbf{Q}_m^{(t)}$, and $\mathbf{X}^{(t)}$, update $\mathbf{Q}^{(t+1)}$ with~\eqref{tilde_solution}.
\STATE   Given $\mathbf{Q}^{(t+1)}$, $\mathbf{T}_m^{(t)}$, $\mathbf{Q}_m^{(t)}$, and $\mathbf{X}^{(t)}$, update $\mathbf{T}^{(t+1)}$ with~\eqref{opt_T}. 
\STATE   Given $\mathbf{Q}^{(t+1)}$, $\mathbf{T}^{(t+1)}$,  and $\mathbf{X}^{(t)}$, update $\mathbf{Q}_m^{(t+1)}$ and  $\mathbf{T}_m^{(t+1)}$ with~\eqref{opt_Qm} and~\eqref{opt_Tm}, respectively. 
\STATE Given $\mathbf{Q}^{(t+1)}$, $\mathbf{T}^{(t+1)}$, $\mathbf{T}_m^{(t+1)}$, and $\mathbf{Q}_m^{(t+1)}$, update $\mathbf{X}^{(t+1)}$  with element-wise optimization.
\STATE Set the iteration index $t=t+1$.
\UNTIL the objective function value of problem~\eqref{problem:sum_rate_X_Re_penalty} experiences a fractional increase smaller than~$\epsilon_1$.
\STATE Reduce the penalty factor with $\tau = \omega\tau$. 
\UNTIL the constraint violation is below threshold $\epsilon_2$.
\end{algorithmic}
\end{algorithm}

\subsubsection{Optimizing $\mathbf{X}$ }
With given $\big\{\{\mathbf{Q}^m\}_{m=1}^{M},\{\mathbf{T}^m\}_{m=1}^{M}\big\}$,  $\mathbf{Q}$, and $\mathbf{T}$, the optimization problem with respect to $\mathbf{X}$ can be expressed as
\begin{subequations}\label{problem:sum_rate_X_Re_penalty_X}
    \begin{align}        
        \min_{\mathbf{X}} \quad  & \!\sum_{m=1}^M\!\left\|\mathbf{T}^m\!-\!\mathbf{H}^m\right\|_F^2 \!+\! \sum_{m=1}^M\!\left\|\mathbf{Q}^m\!-\!\mathbf{G}^m\right\|_F^2 \\
        \mathrm{s.t.}  \quad   & \eqref{constraint:PASS}. 
    \end{align}
\end{subequations}
Adopt the element-wise optimization approach, the optimization problem with respect to $x_n^m$ can be expressed as
\begin{subequations}\label{problem:sum_rate_X_Re_penalty_X_mn}
    \begin{align}        
        \min_{{x}_n^m} \  &    \sum_{k=1}^K\!\left|[\mathbf{T}^m]_{nk}\!-\![\mathbf{H}^m]_{nk}\right|^2 \!\!+\!\! \sum_{p=1}^P\!\left|[\mathbf{Q}^m]_{np}\!-\![\mathbf{G}^m]_{np}\right|^2 \\
          \mathrm{s.t.} \ 
        &  x_n^m \in \mathcal{X}  , \forall m,n.
    \end{align}
\end{subequations}
Problem~\eqref{problem:sum_rate_X_Re_penalty_X_mn} can be solved via a one-dimensional search, where the feasible domain of ${x}_m^n$ is uniformly discretized into~$Q_1$ fine grids.  Problem~\eqref{problem:sum_rate_X_Re_penalty_X} is addressed by iteratively performing the one-dimensional search with respect to $x_n^m, \forall m,n,$  until convergence. 
\begin{remark}
\emph{Under the system setting with one PA per waveguide, the objective function of problem~\eqref{problem:sum_rate_X_Re_penalty_X} can be reformulated as}
\begin{equation}
        \sum_{n=1}^N\!\big(\sum_{k=1}^K\!\left|[\mathbf{T}^1]_{nk}-[\mathbf{H}^1]_{nk}\right|^2 \!+\! \sum_{p=1}^P\!\left|[\mathbf{Q}^1]_{np}-[\mathbf{G}^1]_{np}\right|^2\big).
\end{equation}
\emph{This reformulation reveals that the optimization with respect to $\mathbf{X}$ can be decomposed \emph{in parallel} into $N$ independent single-variable optimization problems. Problem~\eqref{problem:sum_rate_X_Re_penalty_X} can be addressed by solving the resulting $N$ independent single-variable optimization problems with one-dimensional search, respectively.}
\end{remark}
\subsection{The Overall Algorithm}
The proposed AO algorithm for solving problem~\eqref{problem:sum_rate} is summarized in \textbf{Algorithm~2}, where the auxiliary variables and digital beamforming design are iteratively optimized in Section~\ref{digital}, while the pinching beamforming design is addressed in Section~\ref{pinching} using the penalty-based method outlined in \textbf{Algorithm~1}.

\begin{algorithm}[t]
\caption{Overall AO algorithm for solving problem~\eqref{problem:sum_rate_reformulated}.}\label{algorithm3}
\begin{algorithmic}[1]
\STATE {Initialize auxiliary variables $\eta_k^{(0)}$ for all $k$, digital beamforming design $\mathbf{W}^{(0)}$, PA deployment variable $\mathbf{X}^{(0)}$, and the convergence tolerance $\epsilon_3 > 0$. Set the iteration index  $t=0$.}
\REPEAT
\STATE Given $\mathbf{W}^{(t)}$ and $\mathbf{X}^{(t)}$, update $\alpha_k^{(t+1)}$ with~\eqref{alpha_opt_k}. 
\STATE Given $\alpha_k^{(t+1)}$, $\mathbf{W}^{(t)}$, and $\mathbf{X}^{(t)}$, update $\eta_k^{(t+1)}$ with~\eqref{eta_opt_k}. 
\STATE Given $\alpha_k^{(t+1)}$, $\eta_k^{(t+1)}$, and $\mathbf{X}^{(t)}$, update $\mathbf{W}^{(t+1)}$ by solving problem~\eqref{problem:sum_rate_W}. 
\STATE Given $\alpha_k^{(t+1)}$, $\eta_k^{(t+1)}$, and $\mathbf{W}^{(t+1)}$, update $\mathbf{X}^{(t+1)}$ using \textbf{Algorithm~\ref{algorithm4}}. 
\STATE Set the iteration index $t=t+1$.
\UNTIL the objective function value of problem~\eqref{problem:sum_rate_reformulated} experiences a fractional increase smaller than~$\epsilon_3$.
\end{algorithmic}
\end{algorithm}
\begin{figure*}[!h]
\centering
\subfigure[The secondary sum rate versus the power budget of the secondary BS.]{\label{Power_P}
\includegraphics[width=2.3in]{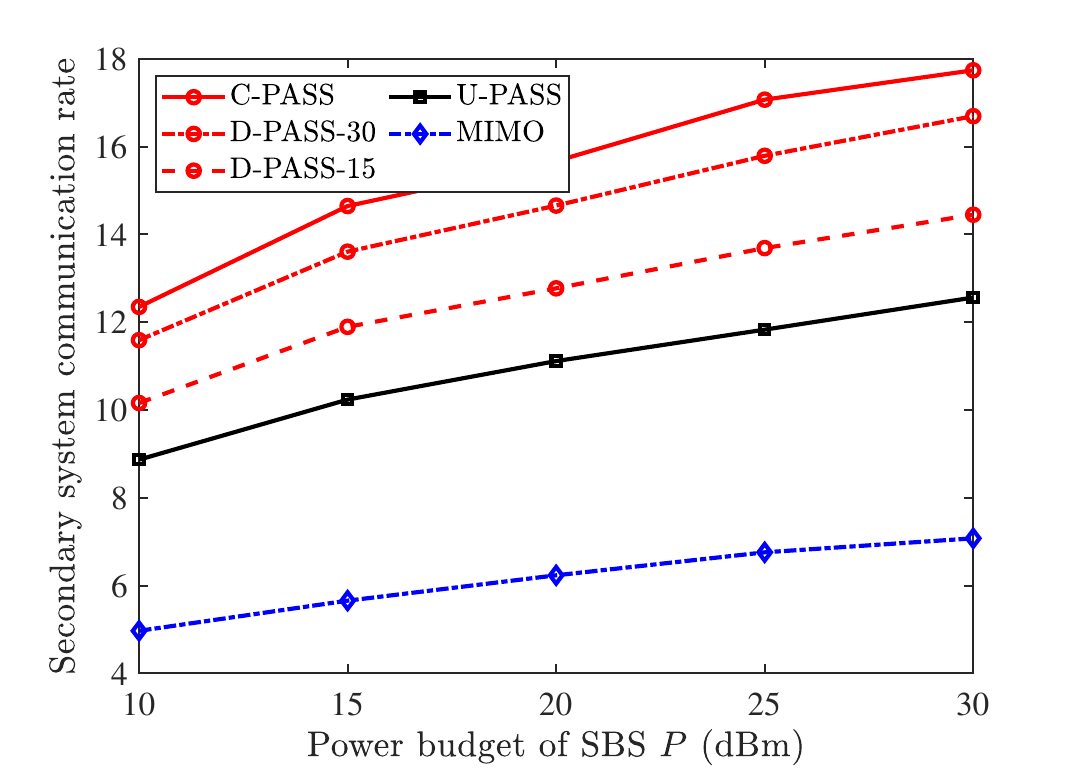}}
\subfigure[The secondary sum rate versus the number of waveguides of the secondary BS.]{\label{Waveguide_N}
\includegraphics[width=2.3in]{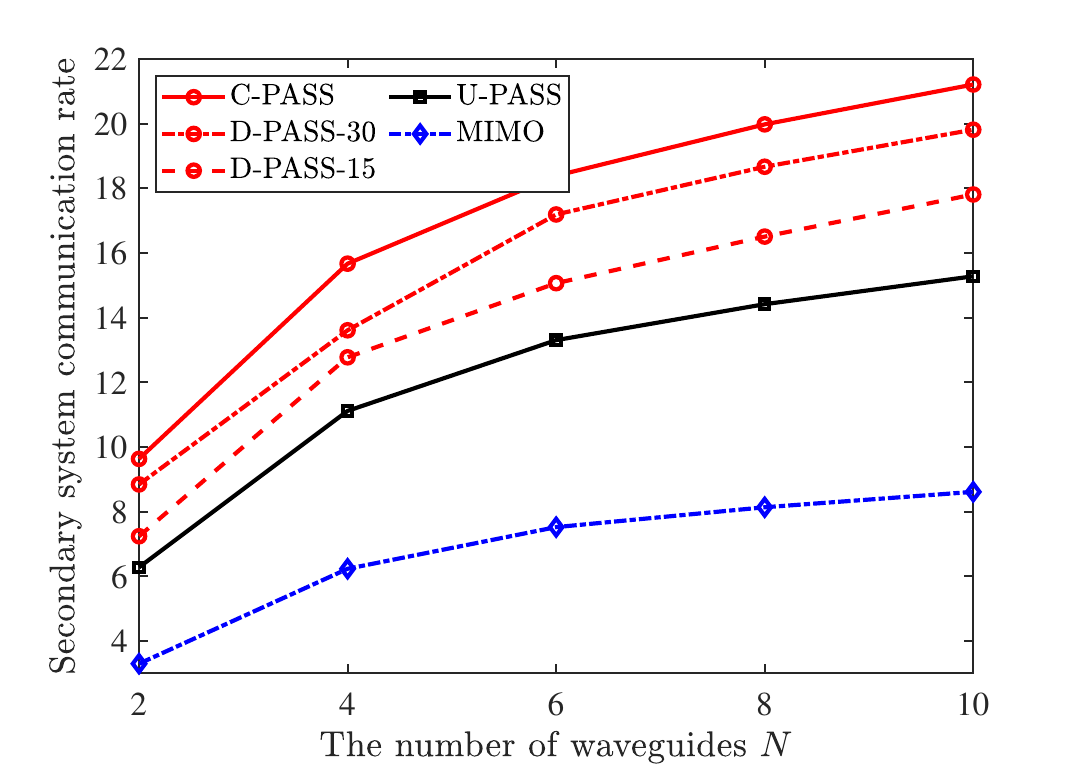}}
\subfigure[The secondary sum rate versus the the IT threshold of primary users.]{\label{IT_Gamma}
\includegraphics[width=2.3in]{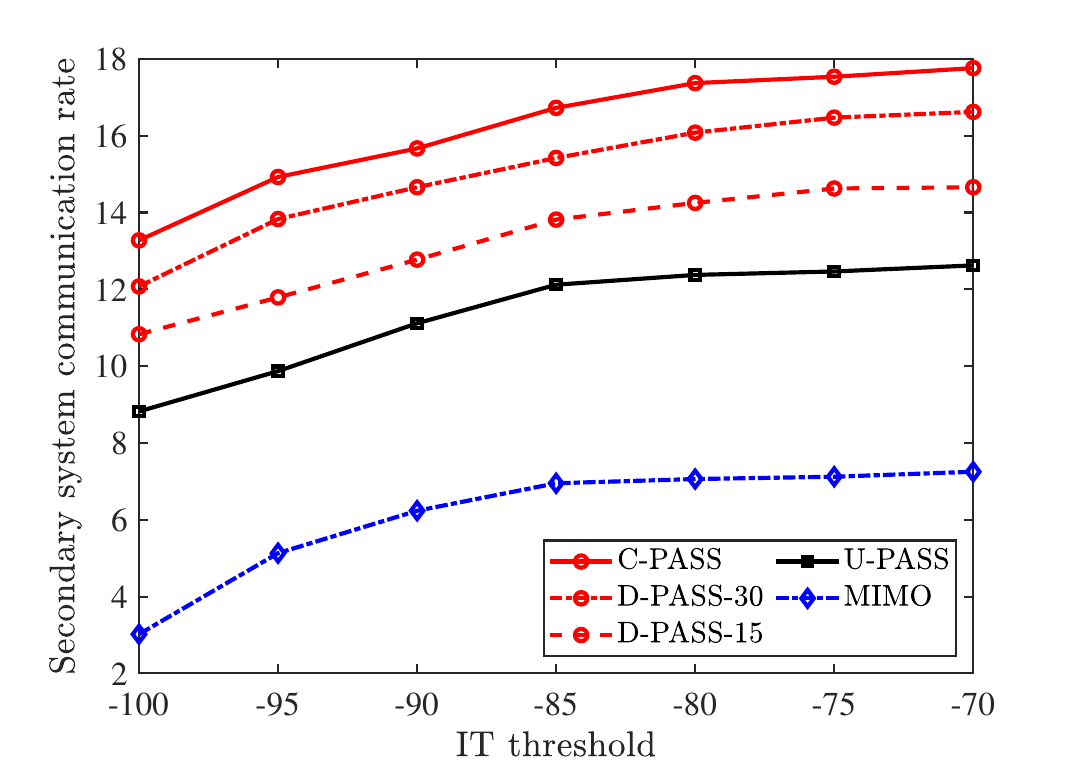}}
\caption{The secondary sum rate versus power budget $P$, number of waveguides $N$, and IT threshold $\Gamma$.}\label{fig:RCRB_UP}\vspace{-0.3cm}
\end{figure*}
\section{Simulation Results}\label{results}
The secondary BS equipped with $N=4$ waveguides and $M=4$ PAs per waveguide is deployed at the coordinate $(0,0,0)$ m. Primary and secondary users are uniformly distributed within the service area on the $X-Y$ plane, which spans $D_x \times D_y = 15\ \text{m} \times 30\ \text{m}$. The number of primary users is $P=2$.  All waveguides are deployed parallel to the x-axis with uniform spacing at a height of $d_h = 3$ m. The length of waveguide is $L=15$ m. The carrier frequency is set as $f_c=28$ GHz. The power limitation at the BS is $P_b=20$ dBm. The noise power at all users is $\sigma^2=-90$ dBm. The IT threshold for primary user $p$ is set as $\gamma_p=\gamma=-90$ dBm. The waveguide
effective refractive index is $n_e=1.4$. Unless otherwise specified, the parameters adopted in the subsequent simulations are consistent with those mentioned above.

Following benchmarks are considered in simulations:
\begin{enumerate}
        \item \textbf{MIMO Scheme}:  In this baseline, the BS employs a ULA and the resulting optimization problem can also be solved using the proposed algorithms in this work by fixing the channels.
        \item \textbf{Uniform PA Scheme}: In this baseline, the PA deployment pattern is not optimized; instead, all PAs are uniformly placed on each waveguide.
        \item \textbf{Discrete PA Scheme}: In this baseline, PAs are restricted to be mounted on $Z$ pre-defined sockets uniformly distributed along the waveguide. The resulting optimization can be addressed using the proposed algorithms.
\end{enumerate}

Fig.~\ref{Power_P} illustrates the sum rate of secondary system versus the power budget of the secondary BS. It can be observed that the sum rate of all considered schemes increases with the power budget. This is because a higher transmit power enables more flexible resource allocation and stronger signal transmission. The continuous PASS enabled CR scheme achieves the best performance due to its highest design flexibility in pinching beamforming. In contrast, the discrete PASS schemes suffer from a reduced beamforming resolution, which limits the available degrees of freedom and leads to performance degradation. Moreover, a coarser discretization results in a more pronounced performance loss. For the uniform PASS scheme, the pinching beamforming capability is completely removed, and thus its performance is further degraded. Nevertheless, it still outperforms the conventional MIMO CR system. This is because the in-waveguide propagation in PASS effectively mitigates the signal propagation loss, thereby enhancing the overall transmission efficiency.

Fig.~\ref{Waveguide_N} illustrates the sum rate of secondary system versus the number of waveguides at the secondary BS. The sum rate of all considered schemes increases with the number of waveguides. This behavior is consistent with the trend observed in conventional MIMO systems, where increasing the number of antennas provides higher spatial degrees of freedom and thus enhances system performance.

Fig.~\ref{IT_Gamma} illustrates the sum rate of the secondary system versus the IT threshold at the primary users. When the IT threshold is small, all schemes are tightly constrained by the interference requirement. As the IT threshold increases, this constraint is gradually relaxed, enabling more flexible resource allocation and thereby improving the achievable sum rate. However, when the IT threshold becomes sufficiently large, the interference constraint is no longer dominant, and the system performance is mainly limited by the transmit power budget, leading to a saturation of the sum rate. The conventional MIMO CR system controls interference leakage to primary users through digital beamforming, whereas the PASS enabled CR system can further suppress interference to primary users by effectively reshaping the channel. As a result, the conventional MIMO scheme is more sensitive to the IT threshold than its PASS-enabled counterpart, particularly in the low-IT regime.

\section{Conclusions}\label{conclusions}
In this paper, we studied a PASS enabled CR system and formulated a secondary capacity maximization problem under transmit power, PA deployment, and interference at primary users constraints. By leveraging a WMMSE-based reformulation and an AO framework, the digital beamforming and pinching beamforming  were optimized in a joint manner. Simulation results verified that the proposed design achieves notable performance gains over benchmark schemes, highlighting the potential of PASS for enhancing spectrum sharing efficiency. 

\bibliographystyle{IEEEtran}
\bibliography{myref}

\end{document}